\documentclass[aps,prb,reprint,amsmath,amssymb,showpacs,superscriptaddress,twocolumn]{revtex4-2}
\usepackage{graphicx}
\usepackage{xcolor}
\usepackage{subfigure}
\usepackage{hyperref,hypcap}
\usepackage{braket}
\usepackage{amsmath}
\usepackage{commath}
\usepackage{mathtools}
\usepackage{chemformula}
\usepackage{natbib}
\usepackage{tabularx}
\usepackage{orcidlink}

\begin{document}
	
	\title{Angular dependence of spin-orbit torque in monolayer \ch{Fe3GeTe2}}
	
	\author{Fei Xue\,\orcidlink{0000-0002-1737-2332}} 
	\affiliation{Department of Physics,	University of Alabama at Birmingham, Birmingham, AL 35294, USA}	
	\affiliation{Physical Measurement Laboratory, National Institute of Standards and Technology, Gaithersburg, MD 20899, USA}
	\affiliation{Institute for Research in Electronics and Applied Physics \& Maryland Nanocenter,	University of Maryland, College Park, MD 20742, USA}
	\author{Mark D. Stiles\,\orcidlink{0000-0001-8238-4156}}	
	\affiliation{Physical Measurement Laboratory, National Institute of Standards and Technology, Gaithersburg, MD 20899, USA}	
	\author{Paul M. Haney\,\orcidlink{0000-0001-9390-4727}}	
	\affiliation{Physical Measurement Laboratory, National Institute of Standards and Technology, Gaithersburg, MD 20899, USA}
	
	\date{\today}
	
	\begin{abstract}
		In ferromagnetic systems lacking inversion symmetry, an applied electric field can control the ferromagnetic order parameters through the spin-orbit torque. The prototypical example is a bilayer heterostructure composed of a ferromagnet and a heavy metal that acts as a spin current source. In addition to such bilayers, spin-orbit coupling can mediate spin-orbit torques in ferromagnets that lack bulk inversion symmetry. A recently discovered example is the two-dimensional monolayer ferromagnet \ch{Fe3GeTe2}. In this paper, we use first-principles calculations to study the spin-orbit torque and ensuing magnetic dynamics in this material. By expanding the torque versus magnetization direction as a series of vector spherical harmonics, we find that higher order terms (up to $\ell=4$) are significant and play important roles in the magnetic dynamics. They give rise to deterministic, magnetic field-free electrical switching of perpendicular magnetization. 
	\end{abstract}
	
	\maketitle
	
	\section{Introduction}
        The electrical control of magnetization without external magnetic fields has attracted a lot of interest due to its potential applications in energy-efficient nonvolatile magnetic random access memory devices and neuromorphic computing~\cite{Manchon2019review,Grollier2020,Shao2021,Hoffmann2022}. One of the promising mechanisms to realize such functionality is spin-orbit torque \cite{miron2011perpendicular,liu2012current,liu2012spin,Manchon2019review,Shao2021}, which is derived from spin-orbit coupling and transfers angular momentum from the crystal lattice to the magnetization~\cite{go2020theory}. 
        The symmetry of the system determines the dependence of the spin-orbit torque on the magnetization direction. This dependence, in turn, determines the possible functionality of the torque in devices. As an example, a bilayer heterostructure consisting of a ferromagnetic and a heavy metal layer often possesses a symmetry mirror plane containing the electric field and the interface normal directions. This symmetry requires that the spin-orbit torque vanishes when the magnetization is in plane and perpendicular to the electric field. This property, in turn, prevents the spin-orbit torque from affecting deterministic switching of magnetic devices with perpendicular magnetic anisotropy, which is desired for applications~\cite{Worledge2011,Bazaliy2015,MacNeill2016,Shao2021,Hoffmann2022}. 
        Utilizing materials with reduced crystal symmetry such as two-dimensional layered materials can overcome this limitation and results in deterministic perpendicular switching~\cite{MacNeill2016,Kao2022,Wang2022Cascadable}. 
        
        In addition to conventional bilayer heterostructures, ferromagnets without inversion symmetry~\cite{Kurebayashi2014} can also exhibit sizable spin-orbit torques, offering another route to useful switching dynamics. An example is the recently discovered 2D magnetic material, monolayer \ch{Fe3GeTe2}. \ch{Fe3GeTe2} is additionally of great interest in ferromagnetic spintronics applications because it is metallic and has strong perpendicular magnetic anisotropy~\cite{Deng2018,Fei2018_FGT,Wang2019current,shi2019}. Johansen {\it et al.} recently predicted that this material's $C_{3z}$ symmetry leads to unique bulk spin-orbit torques~\cite{Brataas2019}.   
        For example, the lowest order spin-orbit torque is found to be time-reversal even and fieldlike, in contrast to the conventional bilayer case that has a time-reversal odd fieldlike torque and a time-reversal even dampinglike torque. Interestingly, although the material symmetry is compatible with deterministic perpendicular magnetization switching, the lowest order torques identified in previous work do not lead to deterministic switching.  
        Motivated by this, we compute the spin-orbit torques in monolayer \ch{Fe3GeTe2} in this paper using {\it ab initio} calculations. We generalize the analysis of the symmetry properties of the material response and show that higher order terms in the spin-orbit torque enable deterministic switching of perpendicular magnetization.
        
        This paper is organized as follows: In Sec.~\ref{sec:sym}, we describe how symmetry determines the form of spin-orbit torques, which we express in vector spherical harmonics. Using vector spherical harmonics as the expansion basis enables the convenient analysis of higher-order terms. We provide symmetry tables for the \ch{Fe3GeTe2} structure and for conventional bilayer systems. Section~\ref{sec:dft} presents first-principles calculations of spin-orbit torques in monolayer \ch{Fe3GeTe2} and analyzes the important higher-order terms in the results. Section~\ref{sec:dynamics} presents the resulting dynamics of the \textit{ab initio} torques computed with the Landau-Lifshitz-Gilbert-Slonczewski equation. 
        In Sec.~\ref{sec:discussion}, we provide a brief discussion of our main findings and relevance to the experiments.

	\section{Symmetry analysis}
	\label{sec:sym}
	\subsection{Vector Spherical Harmonics}
	
	Crystal symmetry ultimately determines the dependence of spin-orbit torque on the electric field and magnetization directions. Following Belashchenko {\it et al.} \cite{Belashchenko2020}, we expand the spin-orbit torque in the basis of vector spherical harmonics. This expansion offers several advantages over other approaches~\cite{Garello2013,Amin2020}  when describing spin-orbit torques in systems with more complicated symmetries than the typical bilayer system. First, the expansion elements are orthogonal to each other, so adding more terms to the expansion does not change the fit values for the lower order terms. Second, there is a straightforward procedure to determine {\it all} symmetry allowed elements of the expansion set. This is in contrast to a polynomial expansion of the torque in Cartesian coordinates, where the number of tensor elements grows exponentially with polynomial order.  This makes higher order terms difficult to identify and evaluate. As we show in this paper, higher order terms (fourth order) can qualitatively impact the features of the spin-orbit torque-induced magnetization dynamics, so their identification is important. Third, the terms in the vector spherical harmonics are automatically partitioned into dampinglike or fieldlike torque terms~\cite{Belashchenko2020}. Knowledge of the fieldlike/dampinglike characteristics of the torque can provide intuition about the role of each term in magnetic dynamics. Finally, the expansion allows easy identification of time-reversal even and odd torques. As we show below, both fieldlike and dampinglike torques include time-reversal even and odd components.  
    We discuss these points in more detail below.

	We follow the same convention adopted in \citet{Belashchenko2020} to use two of the three vector spherical harmonics. For the magnetization direction
    \begin{align}
			\boldsymbol{\hat{m}}=&(\sin\theta \cos\phi,\sin\theta \sin\phi,\cos\theta),
	\end{align}
	the torque components are defined in terms of scalar spherical harmonics $Y_{lm}[\boldsymbol{\hat{m}}(\theta,\phi)]$ as
	\begin{align}
		& \boldsymbol{Y}^{\rm D}_{lm}(\boldsymbol{\hat{m}})
		=\frac{\nabla_{\boldsymbol{\hat{m}}} Y_{lm}(\boldsymbol{\hat{m}})}
		{\sqrt{l(l+1)}},\label{eq:DampingVSH}\\
		&\boldsymbol{Y}^{\rm F}_{lm}(\boldsymbol{\hat{m}})
		=\frac{\boldsymbol{\hat{m}}\times\nabla_{\boldsymbol{\hat{m}}} Y_{lm}(\boldsymbol{\hat{m}})}
		{\sqrt{l(l+1)}},\label{eq:FieldVSH}
	\end{align}
	We explicitly label the vector spherical harmonics terms in Eqs.~\ref{eq:DampingVSH} and Eq.~\ref{eq:FieldVSH} based on the fieldlike or dampinglike nature of the torque. We label $\boldsymbol{Y}^{\rm F}_{lm}$ as fieldlike because its corresponding effective field $\nabla_{\boldsymbol{\hat{m}}} Y_{lm}$ is a pure gradient and has zero curl. Fieldlike torques result in precessional motion of the magnetization. We label $\boldsymbol{Y}^{\rm D}_{lm}$ as dampinglike because it is proportional to $\boldsymbol{m}\times\boldsymbol{Y}^{\rm F}_{lm}$ and can be generated from the curl of an effective field. Dampinglike torques direct the magnetization to fixed points. The time-reversal properties of fieldlike and dampinglike torques depend on whether $l$ is even or odd; Table \ref{table:vsh} summarizes this relationship.

	\begin{table}[htbp]
		\centering
		\begin{tabular}{| c | c| c |}
			\hline
			& $l$ even & $l$ odd \\ \hline
			$\boldsymbol{Y}^{\rm D}_{lm}$ & odd & even \\ \hline
			$\boldsymbol{Y}^{\rm F}_{lm}$ & even & odd \\  \hline  
		\end{tabular}
            \caption{Time-reversal symmetry properties of the vector spherical harmonics $\boldsymbol{Y}^{\rm D,F}_{lm}$.}
            \label{table:vsh}
	\end{table}

     For the most common spin-orbit torques found in bilayers with a broken mirror plane perpendicular to $\bf z$, the dampinglike torque $\hat{\mathbf{m}}\times(({\bf E}\times{\bf \hat z})\times\hat{\mathbf{m}})$ is even under time reversal and the fieldlike torque $\hat{\mathbf{m}}\times({\bf E}\times{\bf \hat z})$ is odd. The terms ``time-reversal even torque'' and ``dampinglike torque'' are often used interchangeably, as are the terms ``time-reversal odd torque'' and ``fieldlike torque''. However, these equivalences do not hold for higher order terms in the expansion of the torque. 

	Since the electric-field-induced spin-orbit torque is always perpendicular to the magnetization $\boldsymbol{m}$ and the vector spherical harmonics form a complete set of functions, we can write  the spin-orbit torkance for an electric field in the ${\boldsymbol{\hat{E}}}$ direction of magnitude $E$
	\begin{equation}
		\boldsymbol{T}_{\boldsymbol{\hat{E}}}(\boldsymbol{\hat{m}})=\boldsymbol{\tau}_{\boldsymbol{\hat{E}}}(\boldsymbol{\hat{m}})E
		\label{eq:tensor}
	\end{equation}
	in the basis of $\boldsymbol{Y}^{\rm D}_{lm}$ and $\boldsymbol{Y}^{\rm F}_{lm}$
	\begin{equation}
		\boldsymbol{\tau}_{\boldsymbol{\hat{E}}}(\boldsymbol{\hat{m}})=\sum_{lm}\left[\boldsymbol{Y}^{\rm D}_{lm}{C}^{\rm D}_{lm}({\boldsymbol{\hat{E}}})+\boldsymbol{Y}^{\rm F}_{lm}{C}^{\rm F}_{lm}({\boldsymbol{\hat{E}}})\right],
	\end{equation}
	where the ${C}$s are complex Cartesian coefficients with the real part being the coefficient of the $\text{Re}\boldsymbol{Y}^{\rm D,F}_{lm} $ and the imaginary parts the negative coefficients of the $\text{Im}\boldsymbol{Y}^{\rm D,F}_{lm} $. The crystal symmetry determines what combinations of coefficients are allowed.
	
When we expand spin-orbit torkance in vector spherical harmonics, we have $2l+1$ independent choices of vector spherical harmonics, one for each integer $m$ with $-l \leq m \leq l$ for a given $l$ in the absence of symmetry constraints. As with spherical harmonics, the vector spherical harmonics with $-m$ are the complex conjugates of those with $m$.
Since the torques are real, we use the real and imaginary parts of the vector spherical harmonics as the expansion functions, e.g., $\text{Re}\boldsymbol{Y}^{\rm D,F}_{lm} $ and $\text{Im}\boldsymbol{Y}^{\rm D,F}_{lm}$.
When we make this choice, we restrict $m$ to be non-negative so we do not overcount. Note that we use a different notation for the vector spherical harmonic torque components than found in Belashchenko {\it et al.}~\footnote{Our vector spherical harmonics convention can be converted to the one adopted by Belashchenko {\it et al.}~\cite{Belashchenko2020}: $\text{Re}\boldsymbol{Y}^{\rm D,F}_{lm}=-\boldsymbol{Z}^{\rm (1),(2)}_{l,m}/\sqrt{2}$, $\text{Im}\boldsymbol{Y}^{\rm D,F}_{lm}=-\boldsymbol{Z}^{\rm (1),(2)}_{l,-m}/\sqrt{2}$}. Crystal symmetries constrain the choices of $m$ for a given $l$. Table \ref{table:symmetryEx} gives the constraints due to important mirror plane symmetries of the structure. Rotational crystal symmetries place additional constraints on $m$, as described in Appendix A.

Conventionally, for thin film heterostructures composed of ferromagnets and heavy metals, the structure is assumed to be disordered, so crystal symmetry does not play a role. The bilayer structure itself breaks the mirror plane $\sigma_{{\hat p},{\hat E}}$, but the other two structural mirror planes remain. The presumed continuous rotational symmetry restricts $m=1$~\cite{Garello2013,Belashchenko2020}, so that for $l$ odd, only \text{Im}$\boldsymbol{Y}^{\rm D,F}_{l1}$ is allowed and for $l$ even, only \text{Re}$\boldsymbol{Y}^{\rm D,F}_{l1}$. 

The material of interest in this paper, \ch{Fe3GeTe2}, preserves the mirror plane perpendicular to the interface normal but breaks one of the mirror planes that contain the interface normal. The mirror plane perpendicular to the interface normal restricts $m$ to be even. When the crystal is orientated such that the electric field is along the $x$ direction as in Fig.~\ref{Fig1:crystal and bands}(a), $\sigma_{{\hat p},{\hat n}}$ is preserved, so terms containing \text{Re}$\boldsymbol{Y}^{\rm D,F}_{lm}$ require $l$ to be odd and terms containing \text{Im}$\boldsymbol{Y}^{\rm D,F}_{lm}$ require $l$ to be even. If the crystal is oriented so that the electric field is along the $y$ direction as in Fig.~\ref{Fig1:crystal and bands}(a), the allowed $l$ values for the different terms switches.
 
Systems like that in Ref.~\cite{MacNeill2016} are similar but do not have the mirror plane perpendicular to the interface, so there is no restriction that $m$ be even. Depending on the orientation of the electric field along the crystal, different terms are allowed for different combinations of $l$ and $m$.

        \begin{table}[htbp]
		\centering

    \begin{tabular}{| c | c| c |}
			\hline
               Mirror plane &  \text{Re}$\boldsymbol{Y}^{\rm D,F}_{lm}$ & \text{Im}$\boldsymbol{Y}^{\rm D,F}_{lm}$ \\ \hline
			$\sigma_{{\hat E},{\hat n}}$ & $l$ even & $l$ odd \\  \hline 			
			$\sigma_{{\hat p},{\hat n}}$ & $l+m$ odd & $l+m$ even \\ \hline
             $\sigma_{{\hat p},{\hat E}}$ & $m$ even  & $m$ even \\ \hline
		\end{tabular}
             \caption{Symmetry constraints on $m$ for a given $l$ provided by different mirror planes in magnetic thin films and thin film heterostructures. The applied electric field is in $\boldsymbol{\hat{E}}$ direction, the film normal direction is $\boldsymbol{\hat{n}}$, and $\boldsymbol{\hat p}=\boldsymbol{\hat n}\times\boldsymbol{\hat E}$. If a system has more than one mirror, the constraints combine. In this table, the vector spherical harmonics are defined based on a coordinate frame such that $\boldsymbol{\hat x}= \boldsymbol{\hat E}$, $\boldsymbol{\hat z}= \boldsymbol{\hat n}$, and $\boldsymbol{\hat y}= \boldsymbol{\hat p}$}
 
            \label{table:symmetryEx}
	\end{table}

It can be informative to take a different approach from that used in Table~\ref{table:symmetryEx}, in which the vector spherical harmonics are defined with respect to the interface normal and the electric field direction and instead to fix the crystal orientation. Then the vector spherical harmonics do not change as the electric field direction is changed  and it becomes possible to relate the coefficients of the different terms for the different electric field directions. This process is explained in Appendix A, allowing us to determine the angular dependence of the torque when the electric field is along $y$ from calculations done for the field along $x$.

    \subsection{General form of the torkance for monolayer \ch{Fe3GeTe2}}

\begin{figure}[htbp]
		\includegraphics[width=1.\columnwidth]{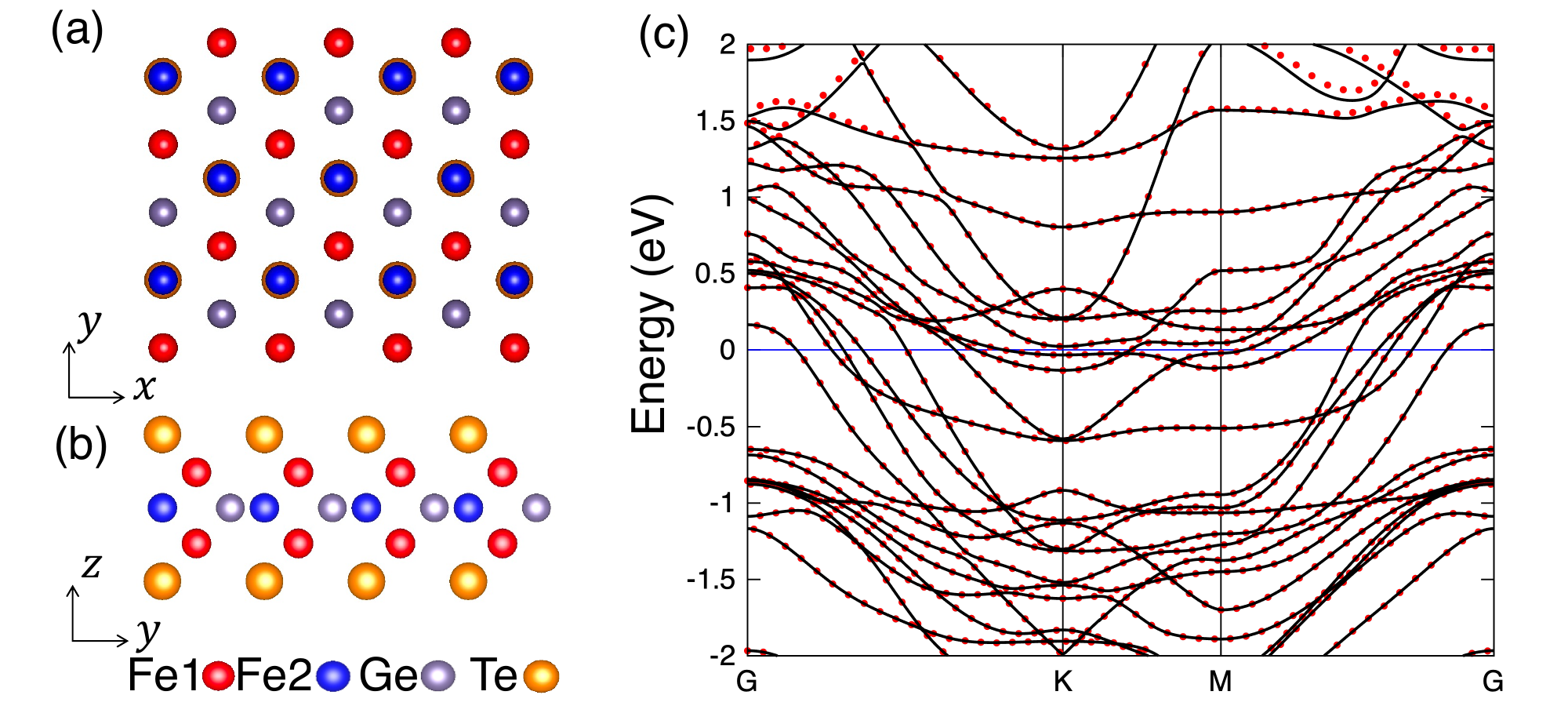}
		\caption{ Crystal structure and band structure of the monolayer \ch{Fe3GeTe2}. (a) Top-down view; (b) side view. We can see that the monolayer \ch{Fe3GeTe2} does not have the mirror symmetry with respect to $xz$-plane. (c) Band structure along the high-symmetry direction G$(0,0,0)$-K$(\frac{1}{3},\frac{1}{3},0)$-M$(\frac{1}{2},0,0)$-G$(0,0,0)$. Red dots represent bands obtained from symmetrized tight-binding Hamiltonian while black lines represent the bands obtained from plane-wave basis. The blue horizontal line indicates the Fermi level.
		}
		\label{Fig1:crystal and bands}
	\end{figure}

	The vector spherical harmonic expansion of the spin-orbit torque for \ch{Fe3GeTe2} is determined by its crystal structure, shown in Fig.~\ref{Fig1:crystal and bands}. Monolayer \ch{Fe3GeTe2} has the $D_{3h}$ symmetry of the $P63/mmc$ space group, which means that it has mirror plane symmetry with respect to the plane of the film ($x-y$ plane), three-fold rotational symmetry around the out-of-plane axis, and three in-plane mirror planes ($y-z$ plane and equivalents rotated by $120^\circ$), but mirror-plane symmetry is broken in the mutually perpendicular planes ($x-z$ plane and equivalents rotated by $120^\circ$). Its lack of inversion symmetry is key to allowing current-induced spin-orbit torque.

\begin{figure*}[htbp]		\includegraphics[width=2\columnwidth]{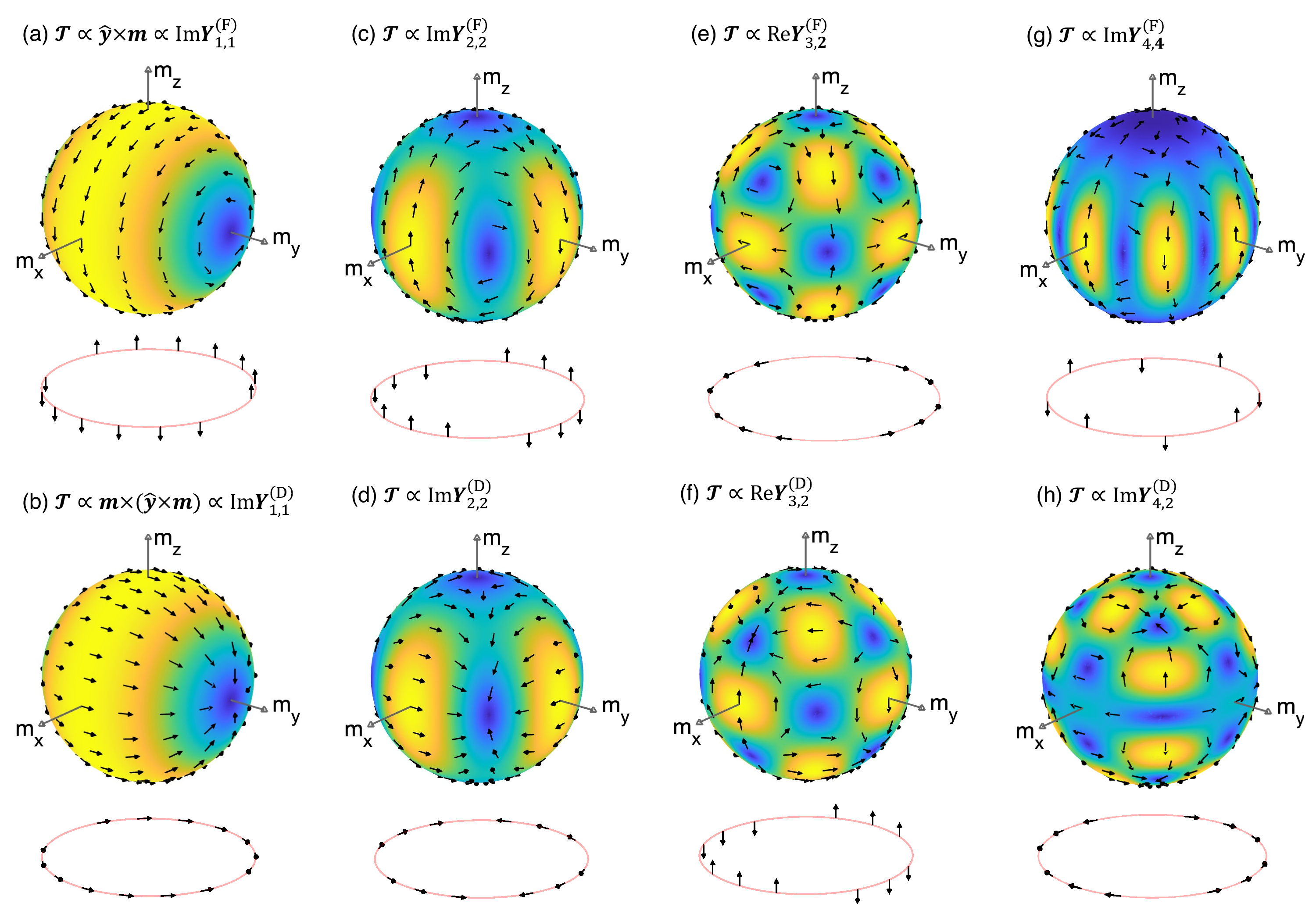}
	\caption{Angular dependence of the exemplary spin-orbit torques for an applied electric field in the $x$ direction. The arrow (color) on the sphere indicates the direction (relative magnitude) of the torque at the given magnetization. The circles below the spheres show the torque along the equator correspondingly. (a) and (b) apply to bilayer heterostructures, while (b)-(h) apply to \ch{Fe3GeTe2}.  Deterministic switching of \ch{Fe3GeTe2} requires torque contributions as shown in (f) and (g).}
	\label{Fig:torque_models}
	\end{figure*}

	Following the general procedure outlined in Appendix~\ref{app:symmetry}, the symmetry-allowed spin-orbit torkance for an electric field in the $\bf x$ direction is given by
 	\begin{equation}
		\begin{split}
			\boldsymbol{\tau}^{\text{even}}_{{\bf \hat{x}}}(\boldsymbol{\hat{m}})=\sum_{lm}&C^{{\rm F}}_{2l,6m\pm2}~\text{Im}\boldsymbol{Y}^{\rm F}_{2l,6m\pm2}(\boldsymbol{\hat{m}})\\
			+&C^{{\rm D}}_{2l+1,6m\pm2}~\text{Re}\boldsymbol{Y}^{\rm D}_{2l+1,6m\pm2}(\boldsymbol{\hat{m}}).
		\end{split} 
		\label{eq:taueven_x_full}
	\end{equation}
	We then project the first-principles results of spin-orbit torkance onto this symmetry-constrained form (up to $\ell=16$) to obtain the full expansion coefficients. Details about the important terms in this expansion are given in Sec.~\ref{sec:dft}.
 	Some of the terms are illustrated in Fig.~\ref{Fig:torque_models}. The lowest order time-reversal even term can be written in Cartesian coordinates as
	\begin{equation}
		\label{eq:fieldlike}
		\begin{split}
			\text{Im}\boldsymbol{Y}^{\rm F}_{2,2}&\propto-\sin{\theta}\cos{2\phi}~\boldsymbol{\hat{\theta}}+\frac{1}{2}\sin{2\theta}\sin{2\phi}~\boldsymbol{\hat{\phi}}\\
			&=\boldsymbol{\hat{m}}\times(m_y,m_x,0) .
		\end{split}
	\end{equation}
	This form, which is shown in Fig.~\ref{Fig:torque_models}(c) and which has been derived from the Cartesian expansion~\cite{Brataas2019}, acts as a fieldlike torque even though it is the time-reversal even component of the spin-orbit torque.

	The time-reversal odd torkance is given by
	\begin{equation}
		\begin{split}
			\boldsymbol{\tau}^{\text{odd}}_{{\bf \hat{x}}}(\boldsymbol{\hat{m}})=\sum_{lm}&C^{{\rm D}}_{2l,6m\pm2}\text{Im}\boldsymbol{Y}^{\rm D}_{2l,6m\pm2}\\
			+&C^{{\rm F}}_{2l+1,6m\pm2}\text{Re}\boldsymbol{Y}^{\rm F}_{2l+1,6m\pm2}.
		\end{split} 
		\label{eq:tauodd_x_full}
	\end{equation}
	The leading term in this expression is in Fig.~\ref{Fig:torque_models}(d), and in Cartesian coordinates takes the form:
	\begin{equation}
		\label{eq:dampinglike}
		\begin{split}
			\text{Im}\boldsymbol{Y}^{\rm D}_{2,2}&\propto\frac{1}{2}\sin{2\theta}\sin{2\phi}~\boldsymbol{\hat{\theta}}+\sin{\theta}\cos{2\phi}\boldsymbol~\boldsymbol{\hat{\phi}}\\
			&=\boldsymbol{m}\times\big((m_y,m_x,0)\times\boldsymbol{m}\big) .
		\end{split}
	\end{equation}
	This time-reversal odd torque acts as dampinglike and is the second lowest-order in magnetization $\boldsymbol{m}$.

	Utilizing Eq.~\ref{eq:EyandEx}, we can write  the final symmetry-constrained form of torkance under the applied E field in the $\boldsymbol{\hat{y}}$-direction by keeping the same coefficients and swapping the $\text{Re}$ and $\text{Im}$ operating on the vector spherical harmonics (see Appendix~\ref{app:symmetry} for details). In this material, even though the coefficients of the torques are the same for fields in the $\boldsymbol{\hat{x}}$ and $\boldsymbol{\hat{y}}$ directions, and the real and imaginary parts of the vector spherical harmonics are the same but rotated through $\pi/m$, the differences between those rotational angles are sufficient to qualitatively change the torques for fields in the two directions. For electric fields in the $\boldsymbol{\hat{y}}$ direction, symmetry prevents magnetic-field free switching of perpendicular magnetizations. However, the different relationship between the electric field and the mirror plane allows for predictable perpendicular switching for an electric field in the $x$ direction. In the following, we focus particularly on this case.

	It is interesting to compare the spin-orbit torques for this system with those typically discussed for bilayer systems. Figures ~\ref{Fig:torque_models}(a) and (b), respectively, show the typical fieldlike and dampinglike torques. These systems have a broken mirror plane perpendicular to the interface normal. When the electric field is applied in plane, both torques vanish when the magnetization points in the in-plane direction perpendicular to the electric field. The torques are finite when the magnetization is perpendicular to the interface. Monolayer \ch{Fe3GeTe2} does not break this mirror plane but rather one containing the interface normal. In this case the torques are strictly zero when magnetizations are perpendicular to the layer. The three fold rotational symmetry then gives more complicated angular dependence than that seen in the bilayer systems. We discuss the consequences of these differences in Sec.~\ref{sec:dynamics}.
	
	A motivation for symmetry analysis is the technological application of current-induced switching of perpendicular magnets \cite{wang2013low,garello2014ultrafast}.  Deterministic spin-orbit torque switching of perpendicular magnetization requires a nonzero out-of-plane torque when the magnetization is along the equator. This form of torque cannot be realized in typical devices composed of isotropic heavy metal layers and ferromagnetic layers due to their in-plane mirror symmetries. The use of in-plane-symmetry-breaking materials such as \ch{WTe2} have been reported previously \cite{MacNeill2016,MacNeill2017,Stiehl2019,Xue2020SOT,Kao2022,Wang2022Cascadable} as a means to accomplishing field-free switching.

    Here we describe a different scenario for achieving deterministic switching of perpendicular magnetizations in \ch{Fe3GeTe2} in which symmetry-allowed higher-order terms in the vector spherical harmonics expansion play an essential role. A first requirement is that when the magnetization is in plane there be an out-of-plane torque to break the symmetry between up and down. Only time-reversal even torques [such as Figs.~\ref{Fig:torque_models}(c), \ref{Fig:torque_models} (f) and \ref{Fig:torque_models}(g)] can provide such functionality because $C_{2y}$ symmetry enforces the out-of-plane torque to have the time-reversal even form, $\tau_z\propto \cos{2m\phi}$. The second requirement is that there will be a stable fixed point out-of-plane, otherwise, the torque will vanish at an in-plane direction. Figure~\ref{Fig:torque_models} (f) shows that torque $\text{Re}\boldsymbol{Y}^{\rm D}_{3,2}$ is the lowest order expansion term to satisfy this requirement. However, a $\rm{Re}\boldsymbol{Y}^{\rm D}_{3,2}$ torque alone cannot switch the magnetization from one hemisphere to the other because of symmetry around the equator for $m=2$ terms. The fixed point in one hemisphere is exactly equivalent to a fixed point at the other hemisphere connected by $(\theta,\phi)\rightarrow{(\pi-\theta,\pi/2-\phi)}$. Although the $\rm{Re}\boldsymbol{Y}^{\rm D}_{3,2}$ torque can drive the magnetization away from the north or south pole when we turn on the field, the new fixed point is still in the same hemisphere. As we turn off the electric field, the magnetization will then go back to the same pole, thus resulting in no switching. The third requirement is breaking the symmetry connecting points in the northern and southern hemispheres which can happen if the higher-order torques with $m>2$ terms are also present. Figure~\ref{Fig:torque_models}(g) shows one example of such torque, $\rm{Im}\boldsymbol{Y}^{(F)}_{4,4}$. The combination of $\rm{Im}\boldsymbol{Y}^{F}_{4,4}$ and $\rm{Re}\boldsymbol{Y}^{D}_{3,2}$ can deterministically switch ferromagnets with perpendicular magnetic anisotropy, as we show in the following sections.

	\section{First-principles calculations of spin-orbit torkances in monolayer \ch{Fe3GeTe2}}
	\label{sec:dft}
	We adopt the experimental unit cell parameters \cite{FGT_unitcell} $a=0.3991~\text{nm}$ of monolayer \ch{Fe3GeTe2} (space group $D_{3h}$) for our first-principles calculations using Quantum ESPRESSO~\cite{QE}. We then use a Wannier function based approach \cite{Wannier90} to compute the linear responses, described in more detail in Appendix~\ref{app:first_principles}. The time-reversal even and odd torkances computed using Kubo formula with the constant broadening and relaxation time approximations are given by ~\cite{Freimuth2014,Xue2020SOT,Xue2021SOT_AFM}
	
	\begin{equation}
		\label{eq:eventorkance}
		\tau^{\rm even}_{ij}=2e\sum_{\substack{\mathbf{k},n,\\
				m\neq n}} f_{n\bf{k}} \frac{\text{Im}\bra{\psi_{n\bf{k}}}  \frac{\partial H_{\bf k}}{\partial k_i}\ket{\psi_{m\bf{k}}}\bra{\psi_{m\bf{k}}}  \mathcal{T}_j\ket{\psi_{n\bf{k}}}}{(E_{m}-E_{n})^2+\eta^2},
	\end{equation}
	\begin{equation}
		\label{eq:oddtorkance}
		\tau_{ij}^{\rm odd}=-e\sum_{\substack{\mathbf{k},n}}\frac{1}{2\eta}\frac{\partial f_{n\bf{k}}}{\partial E_{n\bf{k}}} \bra{\psi_{n\bf{k}}}  \frac{\partial H_{\bf k}}{\partial k_i}\ket{\psi_{n\bf{k}}}\bra{\psi_{n\bf{k}}}  \mathcal{T}_j\ket{\psi_{n\bf{k}}}.
	\end{equation}		
	$\ket{\psi_{n\bf{k}}}$ and $E_{n\bf{k}}$ are the eigenstates and eigenvalues of Hamiltonian $H_{\bf k}$, where ${\bf k}$ is the Bloch wave vector and $n$ is the band index. The equilibrium Fermi-Dirac distribution function is $f_{n\bf{k}}=(e^{ (E_{n\bf{k}}-\mu)/k_{\rm B}T}+1)^{-1}$, $\mu$ is the Fermi level, $\eta$ is the broadening parameter, and $e$ is the electron charge.
	The torque operator is  $\boldsymbol{\mathcal{T}}=-\frac{\mathrm{i}}{\hbar}[{\mathbf{\Delta}}\cdot\hat{\mathbf{S}} ,\hat{\mathbf{S}}]$.
	$\mathbf{S}$ is the spin operator and $\boldsymbol{\Delta}$ is the time-reversal odd spin-dependent exchange-correlation potential.

\begin{figure}[htbp]
		\includegraphics[width=1.\columnwidth]{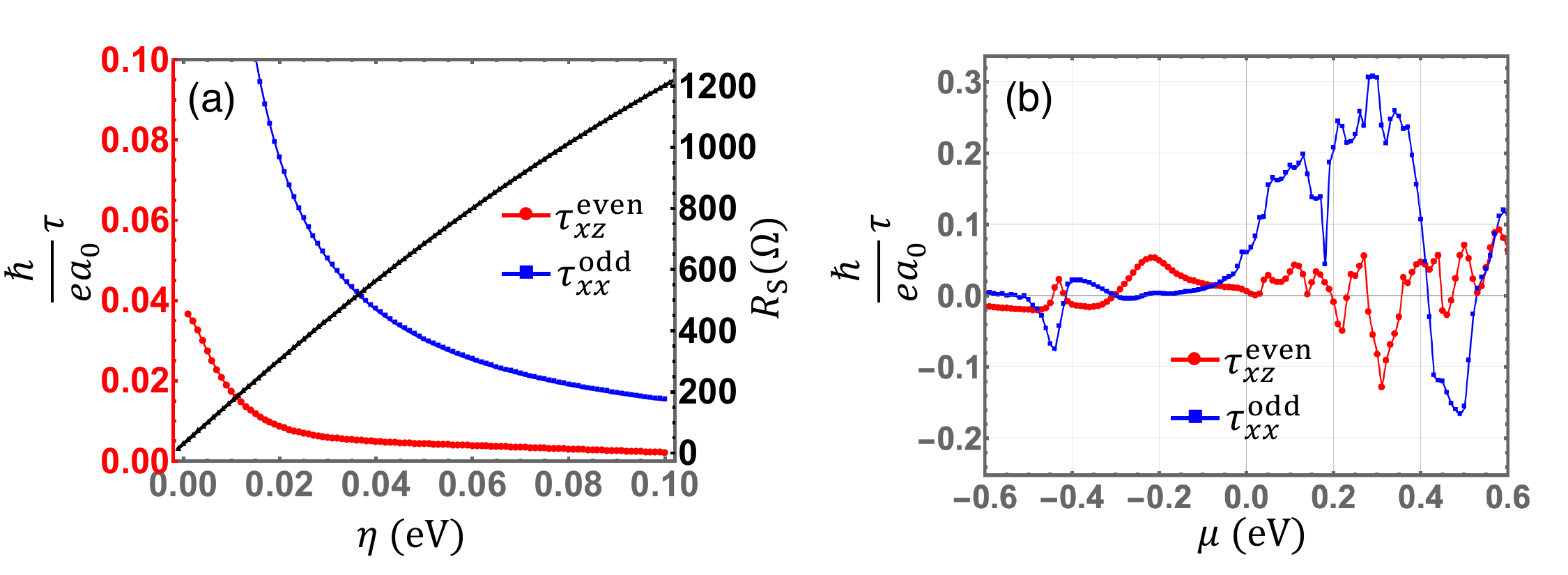}
		\caption{Torkance and sheet resistance as a function of broadening (a) and chemical potential relative to the Fermi level (b). The applied electric field is in the $\hat{{\bf x}}$ direction and the magnetization is in the $\hat{{\bf y}}$ direction. Red and blue lines give the time-reversal even and time-reversal odd torkances respectively. The black line gives the two-dimensional sheet resistance. Note that the time-reversal even torkance is only in the $\hat{\bf{z}}$ direction and time-reversal odd torkance is only in the $\hat{\bf{x}}$ direction due to symmetry constraints. We use $\eta=25~\text{meV}$ in (b).}
		\label{Fig:smearing and mu dependence}
	\end{figure}

	One important input parameter to the calculation is the broadening parameter. 	Figure~\ref{Fig:smearing and mu dependence} shows the dependence of torkance on broadening parameter and chemical potential. In Fig.~\ref{Fig:smearing and mu dependence}(a), we find that the time-reversal odd component $\tau_{xx}$ is always larger than the even component $\tau_{xz}$ when $\boldsymbol{\hat{m}}=\boldsymbol{\hat{y}}$ at the Fermi level. Both time-reversal even and odd torkances increase as the broadening parameter becomes smaller with the odd component increasing faster. The longitudinal resistance is indicated by black line in Fig.~\ref{Fig:smearing and mu dependence}(a). In the broadening parameter regime $\eta\in(0.02,0.04)$~eV, where the resistance is about $400~\Omega$, the odd torkance is almost one order of magnitude larger than the even component. However, the torkance as a function of chemical potential for a fixed $\eta=25~\text{meV}$ shown in Fig.~\ref{Fig:smearing and mu dependence}(b) shows that this ratio does not always hold. Both even and odd components are peaked around $0.3~\text{eV}$ above the Fermi level with a much smaller magnitude difference. In some regions such as $0.2~\text{eV}$ below the Fermi level, the even component can be much larger than the odd component. 
	
	We choose a constant broadening parameter $\eta=25~\rm{meV}$ for the results presented below. The corresponding constant electron momentum relaxation time is $\tau=\hbar/2\eta=13~\text{fs}$. The computed longitudinal resistance [Fig.~\ref{Fig:smearing and mu dependence}(a)] using this $\eta=25~\rm{meV}$ at low temperature is around $400~\rm{\Omega}$ which agrees well with the experiment~\cite{Fei2018_FGT}. Although one experiment \cite{Fei2018_FGT} finds the Curie temperature for monolayer \ch{Fe3GeTe2} can reach up to $100~\rm{K}$, we treat the smaller temperature $T=20~\rm{K}$ \cite{Deng2018} where the ferromagnetic order is most robust.

\begin{figure}[htbp]
		\includegraphics[width=1.\columnwidth]{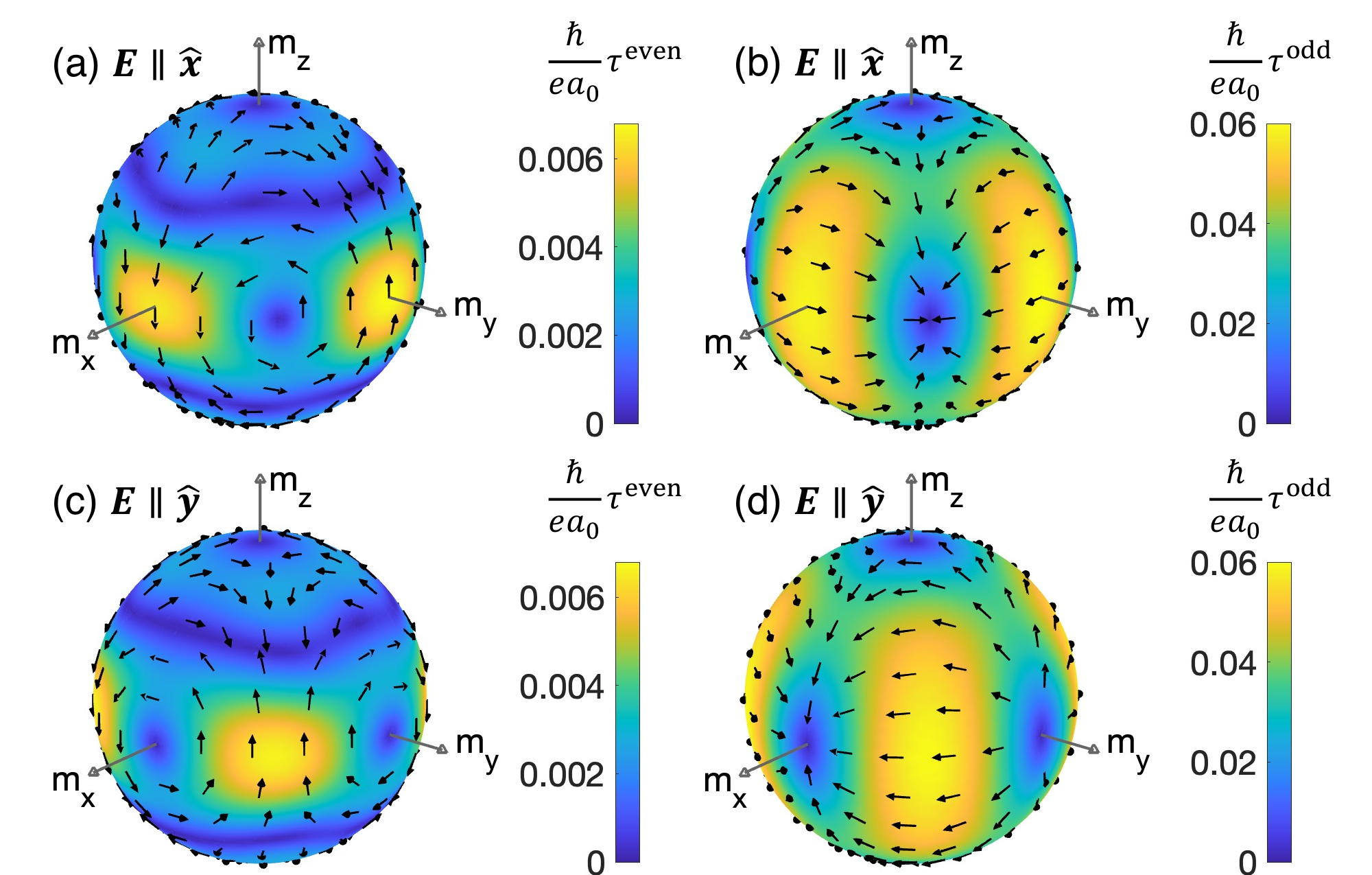}
		\caption{Angular dependence of the time-reversal even (a,c) and time-reversal odd (b,d) torkance on the magnetization direction $(\theta,\phi)$ under an external electric field along the $\boldsymbol{\hat{x}}$ (a,b) direction and $\boldsymbol{\hat{y}}$ (c,d) direction at the Fermi level. The arrow (color) on the sphere indicates the direction (magnitude) of the torkance at the given magnetization. We use $kT=2~\rm{meV}, \eta=25~\rm{meV}$.
		}
		\label{Fig:GlobalTorque}
	\end{figure}

	Figure~\ref{Fig:GlobalTorque} gives the first-principles calculations of spin-orbit torkance in the monolayer \ch{Fe3GeTe2} as a function of magnetization angle $(\theta,\phi)$. 
	Comparing Fig.~\ref{Fig:GlobalTorque}(a) with Fig.~\ref{Fig:torque_models}(c) gives clear evidence of the existence of higher-order terms. There is a vanishing torque band in both north and south hemispheres. 
	We compute the expansion of even and odd torques in vector spherical harmonics as in Eqs.~\ref{eq:taueven_x_full} and \ref{eq:tauodd_x_full} up to $\ell=16$, and find that for $\ell>7$, the coefficients are a factor of 100 smaller than leading order terms. 
 We find that the following three terms in the expansion of the even torque are dominant:
	\begin{equation}
		\begin{split}
			\boldsymbol{\tau}^{\text{even}}_{{\bf \hat{x}}}
			(\boldsymbol{\hat{m}}) ~\approx~ &C^{{\rm F}}_{2,2}\text{Im}\boldsymbol{Y}^{\rm F}_{2,2} +
			C^{{\rm F}}_{4,2}\text{Im}\boldsymbol{Y}^{\rm F}_{4,2}\\
            &+ C^{{\rm D}}_{3,2}\text{Re}\boldsymbol{Y}^{\rm D}_{3,2} .
		\end{split} 
		\label{eq:taueven_Ex}
	\end{equation}
	For the odd torques, the important terms in this expansion are
	\begin{equation}
		\begin{split}
			\boldsymbol{\tau}^{\text{odd}}_{{\bf \hat{x}}}(\boldsymbol{\hat{m}})~ \approx~ &C^{{\rm D}}_{2,2}\text{Im}\boldsymbol{Y}^{\rm D}_{2,2}+
			C^{{\rm D}}_{4,2}\text{Im}\boldsymbol{Y}^{\rm D}_{4,2}\\
			&+C^{{\rm F}}_{3,2}\text{Re}\boldsymbol{Y}^{\rm F}_{3,2}.
		\end{split} 
		\label{eq:tauodd_Ex}
	\end{equation}
The numerical values of these and a few of the next terms in the expansion are given in Tables~\ref{tab:vsh_fit_even} and \ref{tab:vsh_fit_odd}. Using the fitted coefficients of these nonzero vector spherical terms, we can replicate Fig.~\ref{Fig:GlobalTorque}. This allows us to understand specifically how each term contributes to the magnetization dynamics, the focus of the next section. 
 
	\begin{table}[htbp]
		\centering
  
		\begin{tabular}{|c|c|c|c|c|c|}			\hline					
			$C^{{\rm F}}_{2,2}$ &$C^{{\rm F}}_{4,2}$
			&$C^{{\rm F}}_{6,2}$
			&$C^{{\rm F}}_{4,4}$
			&$C^{{\rm F}}_{6,4}$
			&$C^{{\rm D}}_{3,2}$ \\
			\hline
			$-0.0087$ & 0.0096 & $-0.0015$ & 0.0007 &  0.0006 & $-0.0075$ \\ 
			\hline
		\end{tabular}

		\caption{Expansion coefficients of the time-reversal even torques as in Eq.~(\ref{eq:taueven_x_full}). The torques are in units of $ea_0/\hbar$ where $a_0$ is the Bohr radius. Other terms with magnitudes less than 0.0005 are not shown in this Table.}
		\label{tab:vsh_fit_even}
	\end{table}
	\begin{table}[htbp]
		\centering
  
	            \begin{tabular}{|c|c|c|c|}			\hline					
			$C^{{\rm D}}_{2,2}$&
			$C^{{\rm D}}_{4,2}$&
			$C^{{\rm F}}_{3,2}$&
                $C^{{\rm F}}_{7,2}$\\
			\hline
			0.2018 & 0.0017 & 0.0153 & 0.0013  \\ 
			\hline
		\end{tabular}

		\caption{Expansion coefficients of the time-reversal odd torques as in Eq.~(\ref{eq:tauodd_x_full}). The torques are in units of $ea_0/\hbar$ where $a_0$ is the Bohr radius. Other terms with magnitudes less than 0.001 are not shown in this table.}
		\label{tab:vsh_fit_odd}
	\end{table}

	Figures~\ref{Fig:GlobalTorque} (c) and \ref{Fig:GlobalTorque} (d) show the angular dependence of spin-orbit torques when the applied electric field is in $\boldsymbol{\hat{y}}$ direction. Because of the $C_{3z}$ rotation symmetry, these results are expected to be related with the results of the applied field in the $\boldsymbol{\hat{x}}$ direction according to Eq.~\ref{eq:EyandEx}. 
    We have checked that the numerical results are indeed consistent with this relationship. If we look at each individual vector spherical harmonic term, the difference between the cases for $\boldsymbol{E}\parallel\boldsymbol{\hat{y}}$ and $\boldsymbol{E}\parallel\boldsymbol{\hat{x}}$ is a simple azimuthal rotation by an angle of $\frac{\pi}{2m}$ to swap the real and imaginary parts. After summing over all $m$, the total torques for the two cases are not related by a simple rotation. This enables an out-of-equator fixed point for $\boldsymbol{E}\parallel\boldsymbol{\hat{x}}$, as we describe next.

    \begin{figure}[htbp]
	\includegraphics[width=1.\columnwidth]{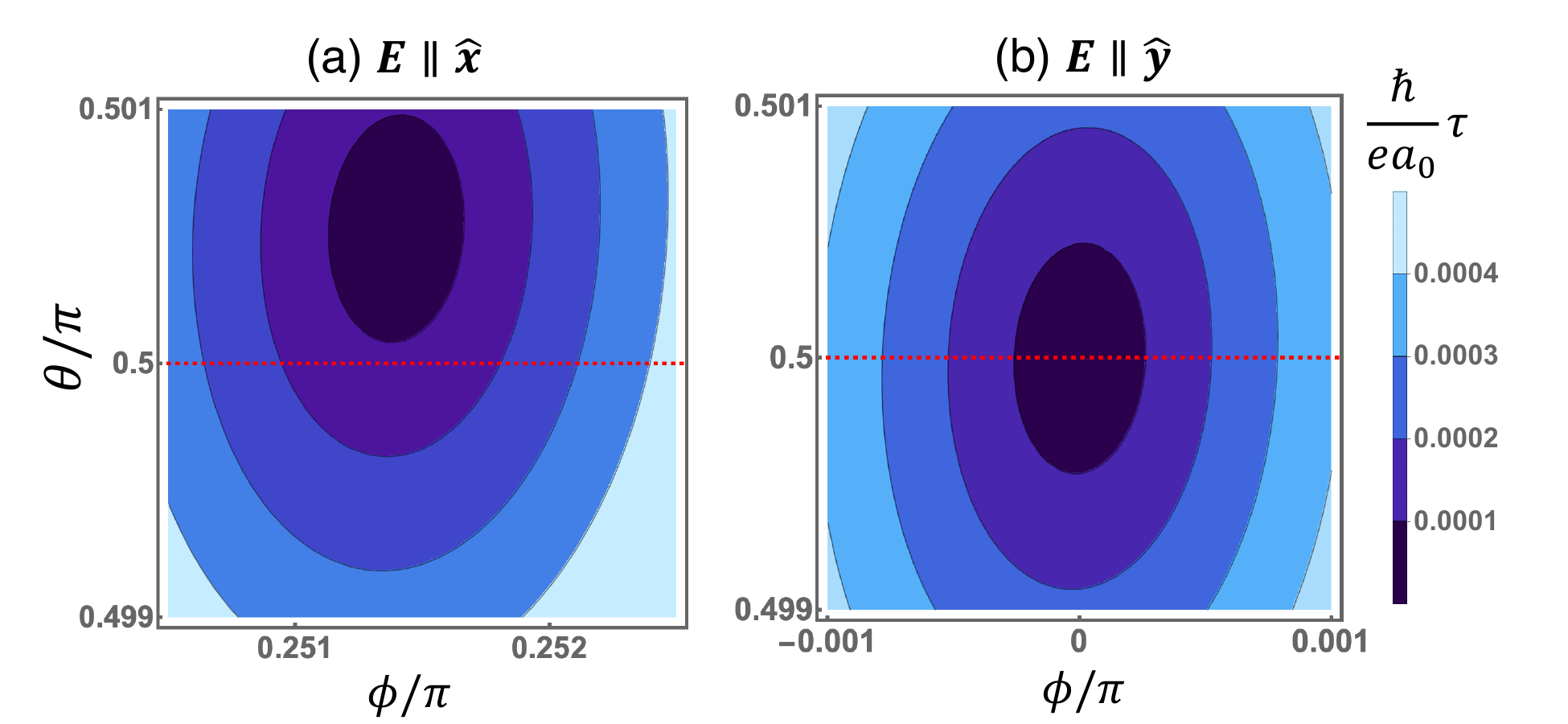}
	\caption{Zoomed-in contour plot of torkance magnitude as a function of magnetization direction $(\theta,\phi)$ under an external electric field along the $\boldsymbol{\hat{x}}$ (a) direction and $\boldsymbol{\hat{y}}$ (b) direction at the Fermi level. Horizontal red dashed line indicates the equator position where $\theta=\pi/2$. (a) Zero torkance near $\phi=\pi/4$; (b) zero torkance at $\phi=0$.
		}
	\label{Fig:contourplot}
    \end{figure}
    Figure~\ref{Fig:contourplot} shows a zoomed-in contour plot of the magnitude of the total spin-orbit torkance near the equator. In the case of $\boldsymbol{E}\parallel\boldsymbol{\hat{y}}$, the mirror symmetry $\sigma_{yz}$ enforces a zero torkance fixed point at $\boldsymbol{m}=\boldsymbol{\hat{x}}$, shown in Fig.~\ref{Fig:contourplot}(b). Microscopically, all vector spherical harmonic terms in Eq.~\ref{eq:tauodd_Ey_full} are zero when $(\theta,\phi)=(\pi/2,0)$. In contrast, Fig.~\ref{Fig:contourplot}(a) shows one of the four out-of-equator zero torkance fixed points near $(\theta,\phi)=(\pi/2,\pi/4)$. The fixed points in (a) and (b) are inequivalent due to the broken $\sigma_{xz}$ mirror symmetry in \ch{Fe3GeTe2}. The three additional zero-torkance points include one on the same hemisphere and two on the opposite hemisphere. For a particular electric field, the two fixed points on the same hemisphere are stable and the other two on the opposite hemisphere are unstable. The stability of each points changes with the sign of the electric field, allowing deterministic switching, discussed in the next section.
    
\begin{figure}[htbp]
		\includegraphics[width=1\columnwidth]{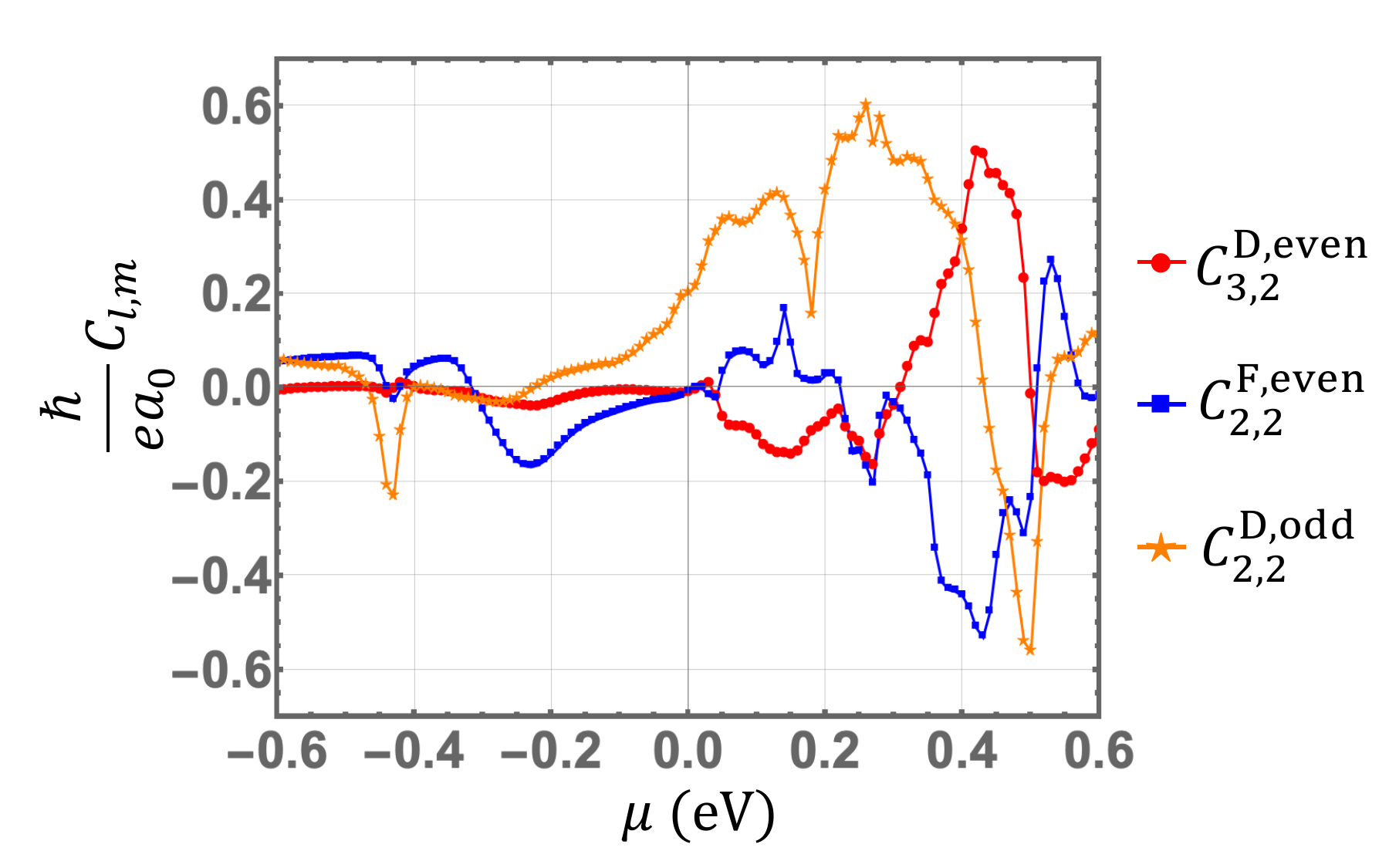}
		\caption{Fitted parameters $C_{l,m}$ as a function of chemical potential relative to the Fermi level. The applied electric field is in $\hat{{\bf x}}$ direction. Red dots represent the $C^{\rm{D,even}}_{3,2}$ coefficients which is crucial in generating out-of-equator fixed points demonstrated in Fig.~\ref{Fig:torque_models}(f). Blue squares and orange stars represent the lowest order time-reversal even and time-reversal odd fitted parameters, corresponding to Fig.~\ref{Fig:torque_models}(c) and (d) respectively.  We use $\eta=25~\text{meV}$.}
		\label{Fig:mu dependence}
    \end{figure}

\begin{figure}[htbp]
	\includegraphics[width=1.\columnwidth]{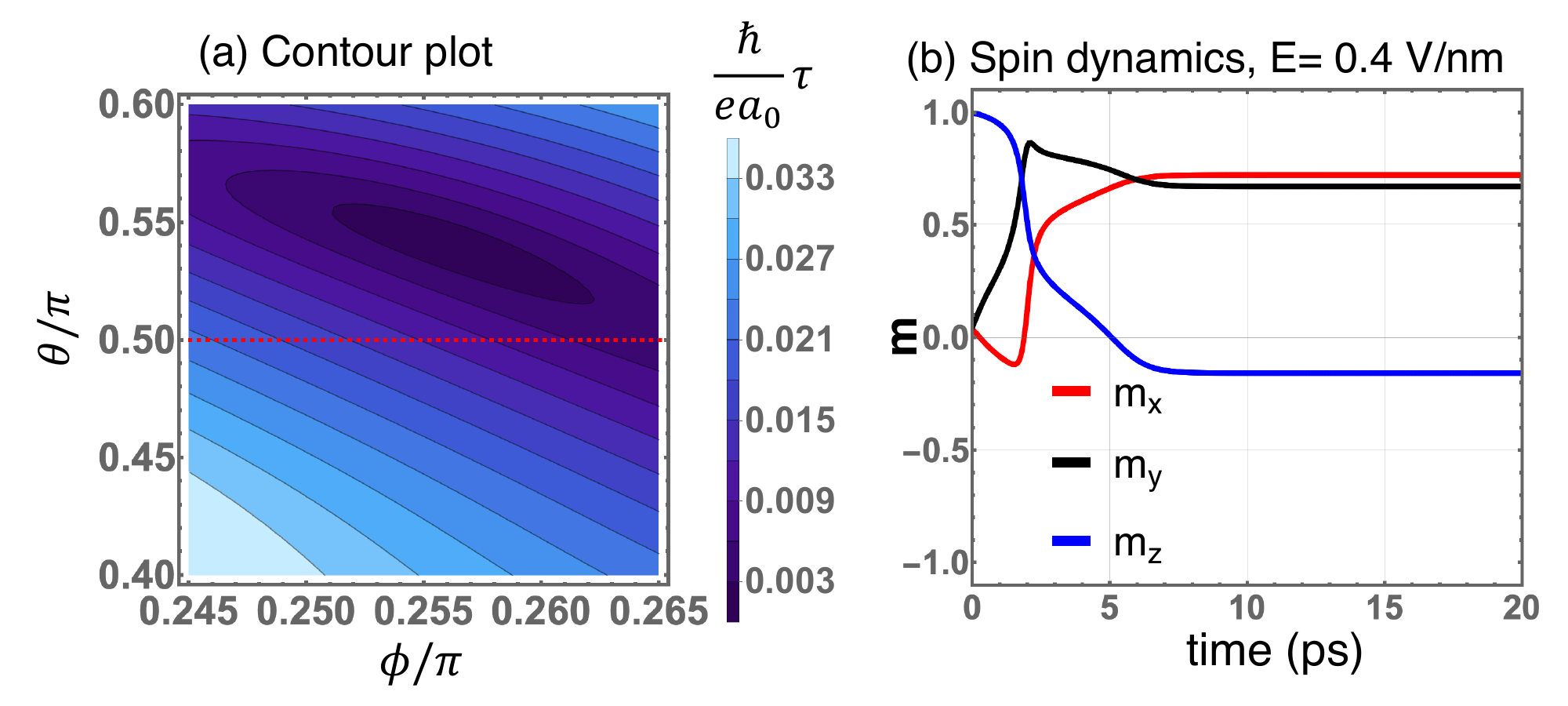}
	   \caption{(a) shows a zoomed-in contour plot of torkance magnitude as a function of magnetization direction $(\theta,\phi)$ under an external electric field along the $\boldsymbol{\hat{x}}$ direction at $\mu=0.1~\text{eV}$ above the Fermi level. Horizontal red dashed line indicates the equator position where $\theta=\pi/2$. (b) shows the spin dynamics in the presence of electric-field-induced ($\boldsymbol{E}\parallel\hat{\boldsymbol{x}}$) \textit{ab initio} spin-orbit torkance. Red, blue, and black lines represent the dynamics of $m_x,m_y,m_z$ respectively. We use $\alpha=0.01$, $\mu_0H_{\rm A}=20~\text{T}$, and $\mu=0.1~\rm{eV}$ in the Landau-Lifshitz-Gilbert dynamics simulation. The applied electric field strength $0.4~\rm{V/nm}$ gives a current density of $4\times10^{14}~\rm{A/m^2}$  assuming a $2.5~\rm{nm}$ sample thickness and a $400~\Omega$ sheet resistance.}
	\label{Fig:mu=0.1eV}
	\end{figure}

    Fig.~\ref{Fig:contourplot}(a) shows a tiny polar angle difference from $\pi/2$ which is unlikely to be thermally stable in realistic applications. The reason the angle is so small is that $C^{\text{D}}_{3,2}$ is relatively small compared to lower order terms such as $C^{\text{D,F}}_{2,2}$ which all have the fixed points at equator. The smallness of $C^{\text{D}}_{3,2}$ is not always true, as shown in the fitted coefficients as function of the chemical potential in Fig.~\ref{Fig:mu dependence}. The important $C^{\text{D}}_{3,2}$ term can be very prominent as we increase chemical potential a few tens of millielectron volts indicated by the red line. At this chemical potential range, the out-of-equator fixed point can be detectable much more easily, as shown in the contour plot of Fig.~\ref{Fig:mu=0.1eV} (a). While the properties we calculate of \ch{Fe3GeTe2} are not likely to be suitable for applications, our focus is on  the new physics and its trends dictated by the symmetries in \ch{Fe3GeTe2}, rather than specific values. Other materials that share the same symmetry may have properties that are more amenable.

	\section{Dynamics}
	\label{sec:dynamics}
	
	In this section, we focus on how the spin-orbit torques computed in the previous section  affect the magnetization dynamics. The spin dynamics of a ferromagnet with perpendicular easy-axis anisotropy is governed by the following Landau-Lifshitz-Gilbert equation with additional current-induced spin-orbit torque terms~\cite{SLONCZEWSKI1996}
	\begin{equation}
		\label{eq:LLG}
		\frac{d\hat{\mathbf{m}}}{dt}-\alpha\hat{\mathbf{m}}\times\frac{d\hat{\mathbf{m}}}{dt}=-\gamma \mu_0H_{\rm A} \left(\hat{\mathbf{m}}\times 	\hat{{\bf z}}\right)\left(\hat{\mathbf{m}}\cdot\hat{{\bf z}}\right)+\boldsymbol{\mathcal{T}},
	\end{equation}
	where $\hat{\mathbf{m}}$ is the normalized magnetization, $\alpha$ is the Gilbert damping parameter, $\gamma$ is the absolute value of the gyromagnetic ratio, $\mu_0$ is vacuum magnetic permeability, $H_{\rm A}$ is the magnetic anisotropy field, and $\mathcal{T}$ is the current-induced spin-orbit torque.

	We directly compute the spin dynamics with the \textit{ab initio} fitted spin-orbit torques as input into the Eq.~\ref{eq:LLG}. We use the full expansion of the torque (up to $\ell=16$) even though we obtain nearly identical results in test calculations significantly truncating the expansion. In the simulation, we choose $\mu_0 H_{\rm A}=20~\rm{T}$ by calculating the energy difference for out-of-plane and in-plane magnetic configuration~\cite{FGT_anisotropy2016}. For the Gilbert damping, we choose $\alpha=0.01$~\cite{FGT_damping2022}. Fig.~\ref{Fig:mu=0.1eV}(b) shows a typical zero-temperature magnetic trajectory when the applied electric field is larger than a critical threshold. The stable fixed point $(\theta_E,\phi_E)$ corresponds to the same fixed point near $\phi=\pi/4$ determined by the spin-orbit torkance shown in Fig.~\ref{Fig:mu=0.1eV}(a) but shifted by the presence of the anisotropy torque. There is another electric-field driven stable point near the symmetry related fixed point $(\theta_E,\phi_E+\pi)$ depending on the initial state of the magnetization. Reversing the sign of electric field makes the other two fixed points [$(\pi-\theta_E,-\phi_E)$ and $(\pi-\theta_E,-\phi_E-\pi)$] become stable so that it is possible to switch the magnetization from the south pole to the north hemisphere.

    The spin-orbit torques in monolayer $\ch{Fe3GeTe2}$ lead to dynamics that are quite distinct from those of the conventional cases as analyzed in Ref.~\cite{Bazaliy2015}. First, the instability condition of the initial magnetization is very different from the cases found in bilayers. In the bilayer case, for a perpendicular easy-axis anisotropy, the spin-orbit torque is finite on the initial magnetization ($\pm\hat{\mathbf{z}}$), see Fig.~\ref{Fig:torque_models}(a,b). For \ch{Fe3GeTe2} on the other hand, the torque on that initial magnetization is zero by symmetry as seen in Fig.~\ref{Fig:GlobalTorque}. For this aspect of the reversal, the initial instability for \ch{Fe3GeTe2} has more in common with the instability for a bilayer system with an in-plane easy-axis along the $\pm\hat{\mathbf{y}}$, because in that case the torque is also zero.
    
    The instability case for \ch{Fe3GeTe2} also differs significantly from that of the bilayer with in-plane easy-axis anisotropy. As seen in Fig.~\ref{Fig:torque_models}(b), when the magnetization in the bilayer system precesses around the easy axis, the dampinglike torque pushes magnetization toward the easy axis or away from it depending on the sign of the current but independent of the phase of the precession. This means that the dampinglike torque competes with the damping torque, which is a factor of $\alpha$ smaller than the precession torques. On the other hand, the torques shown in Fig.~\ref{Fig:GlobalTorque}(a,b) have no net push toward the easy axis along the poles (due to the $\sigma_{xy}$ symmetry making the poles saddle points for the spin-orbit torques) and so they do not compete with damping torque. For \ch{Fe3GeTe2}, when the magnetization is near the poles, the spin-orbit torques compete with the anisotropy directly. This competition gives the unfortunate consequence that reversal instability in \ch{Fe3GeTe2} requires larger currents than might be the case for other symmetries. However, when the magnetization is close to the fixed points near the equator, the spin-orbit torque competes directly with the damping, giving smaller critical currents for the stability of those fixed points.
    
    Once the critical current is reached and the $\hat{\mathbf{z}}$ direction becomes unstable for the magnetization, \ch{Fe3GeTe2} has the advantage over the bilayer system with perpendicular anisotropy that the switching is deterministic without any other symmetry breaking, like in-plane magnetic fields, applied to the system. In the bilayer system without symmetry breaking the magnetization goes to the $\hat{\mathbf{y}}$. When the current is turned off, small fluctuations determine whether the magnetization reverses or returns to its original state. For \ch{Fe3GeTe2} on the other hand, as shown in Fig.~\ref{Fig:mu=0.1eV}, the stable minima near $m_z=0$ are in one equator or the other, so that when the current is turned off, the magnetization goes to the pole on that side of the equator.

	\section{Discussion}
	\label{sec:discussion}
        Our findings have several experimental implications. The lowest order $\text{Im}\boldsymbol{Y}^{\rm F}_{2,2}$ has been found to be important in assisting the conventional dampinglike torque $\text{Im}\boldsymbol{Y}^{\rm D}_{1,1}$ in perpendicular switching of bilayer CoPt/CuPt~\cite{Chen2021,Manchon2023}. This combination shares the similar traits as \ch{Fe3GeTe2}. Reversal requires mixing vector spherical harmonics with different $m$ and nonzero out-of-plane torques when the magnetization is in-plane. Our numerical results also give a large time-reversal odd dampinglike torque $\text{Im}\boldsymbol{Y}^{\rm D}_{2,2}$ in \ch{Fe3GeTe2}, which can be tested in existing second harmonics setups~\cite{Klaui2023}.

        In order to quantify all the symmetry-allowed higher order torques, a complete sweep of magnetization is required. Similar work has been done in \ch{WTe2}/\ch{Ni80Fe20} bilayer~\cite{MacNeill2016}. Instead of expanding the measured torques into trigonometric functions, we need to expand them into vector spherical harmonics and obtain the fitting parameters. As we have shown, the coefficients vary largely as we change the chemical potential. Thus, adding a bias gate to change the charge density~\cite{Deng2018} in monolayer \ch{Fe3GeTe2} might be a way to find useful experimental conditions. 

        The critical electric field to switch the perpendicular magnetization in \ch{Fe3TeGe2} is high because the mirror symmetry $\sigma_{xy}$ restricts  torques to those with even $m$. This restriction requires the spin-orbit torques to compete with the anisotropy torque instead of the damping torque. This mirror symmetry can be broken in the presence of a substrate or applied out-of-plane electric field, similar to the case of bilayer CoPt/CuPt~\cite{Chen2021,Manchon2023}.

        In summary, we perform first-principles calculations of spin-orbit torque in monolayer \ch{Fe3GeTe2} and discover that the bulk spin-orbit torque expressed in higher-order vector spherical harmonics can deterministically switch the perpendicular magnetization. We have provided a symmetry table for other reduced symmetry systems as well. Utilizing higher-order spin-orbit torque offers a new perspective to realize novel electrical control of magnetization.

        \section{Acknowledgement}
        We thank Kirill Belashchenko and Alexey Kovalev for useful discussions. The work done at University of Alabama at Birmingham is supported by the National Science Foundation under Grant No. OIA-2229498, UAB internal startup funds, and UAB Faculty Development Grant Program, Office of the Provost.
        F.X. also acknowledges support under the Cooperative Research Agreement between the University of Maryland and the National Institute of Standards and Technology Physical Measurement Laboratory, Award 70NANB14H209, through the University of Maryland.
	\appendix
	
	\section{Symmetry-constrained form of spin-orbit torque in vector spherical harmonics basis}
	\label{app:symmetry}
	
	The symmetry allowed form of spin-orbit torque tensor can be obtained by averaging all possible symmetry transformed tensors $\boldsymbol{\tau}^{\text{sym}}=\frac{1}{N}\sum\boldsymbol{\tau}'$ where $N$ is the number of symmetry operations and $\boldsymbol{\tau}'$ indicates the tensor after the transformation. If we consider an orthogonal transformation to a Cartesian tensor, we can usually get the explicit transformation form under a rotation $R$
	\begin{equation}
 \tau'_{ijk...}=\sum_{\alpha\beta\gamma...}\det(R)R_{i\alpha}R_{j\beta}R_{k\gamma}...\tau_{\alpha\beta\gamma...}.
		\label{eq:cartesian_expansion}
	\end{equation}
In Cartesian form such as $T_i=\tau_{ijk}E_j m_k$ to arbitrary order, the number of nonzero components in the tensor $\tau$ becomes exponentially large as we increase the tensor rank. It becomes practically intractable to obtain the symmetry-allowed higher-order terms in $\hat{\boldsymbol{m}}$ of $\tau$ in Cartesian form. 

We next describe the transformation of the torkance tensor in the expansion of vector spherical harmonics. For this purpose, it's convenient to write the tensor with a slightly different notation than used in the main text. In what follows, the tensor ${{\boldsymbol \tau}}$ relates the electric field ${\bf E}$ to torque ${\bf T}$ according to: 
\begin{eqnarray}
    {\bf T} &=& {{\boldsymbol \tau}} \cdot {\bf E} 
\end{eqnarray}
 ${{\boldsymbol \tau}}$ is the outer product of a vector spherical harmonic ${\bf Y}$ which specifies the torque direction, and a row vector ${\bf C}$ that contracts with the electric field: 
\begin{eqnarray}
    {{\boldsymbol \tau}} &=& {\bf Y}(\theta,\phi) \otimes {\bf C}
\end{eqnarray}
A coordinate transformation $U$ of the system will act on both electric field and magnetization directions, and is represented by $U_{\hat M}$ and $U_{\hat E}$, respectively:
\begin{eqnarray}
    U_{\hat M} {\bf T} &=& {{\boldsymbol \tau}} \cdot (U_{\hat E} {\bf E})  
\end{eqnarray}
For operations which leave the crystal invariant, we require that the transformed torkance is also invariant, so that ${{\boldsymbol \tau}}$ satisfies:
\begin{eqnarray}
    {{\boldsymbol \tau}} &=& U_{\hat M}^{-1} {{\boldsymbol \tau}}U_{\hat E}
\end{eqnarray}
The above equation provides symmetry constraints on ${{\boldsymbol \tau}}$ for a given symmetry transformation $U$.

In the following, we apply this procedure for \ch{Fe3GeTe2} for each of the materials symmetry operations. The monolayer \ch{Fe3GeTe2} has the point group symmetry $D_{3h}$ \cite{Brataas2019} which consists of one $C_3$ rotations around $z$-axis, three $C_2$ rotations including one around $y$-axis, and one mirror reflection respect to $xy$-plane, as shown in Fig.~\ref{Fig1:crystal and bands}. Since we are interested in the case where electric field is applied in-plane, it is convenient to consider the rotation symmetry around $z$-axis first.
	According to Eq.~\ref{eq:tensor}, the torkance tensor $\boldsymbol{\tau}$ is invariant under a rotation because both torque $\boldsymbol{T}$ and electric field $\boldsymbol{E}$ follow the same transformation under a rotation. Since the vector spherical harmonics absorbs an extra phase under a rotation of angle $\gamma$, i.e., ${\boldsymbol{Y}}^{(\nu)}_{lm}(\theta,\phi-\gamma)\rightarrow {\boldsymbol{Y}}^{(\nu)}_{lm}(\theta,\phi)e^{-im\gamma}$, the transformed vector coefficients $\boldsymbol{C}$ need to have additional phase factors $e^{im\gamma}$ to compensate $e^{-im\gamma}$ in order to keep the tensor $\boldsymbol{\tau}$ invariant.
	If the rotation symmetry is continuous, the only possible way is either $m=0, \boldsymbol{C}\propto \boldsymbol{\hat{{z}}}$ or $m=\pm1, \boldsymbol{C}\propto \boldsymbol{\hat{{x}}}\pm i\boldsymbol{\hat{{y}}}$. We can then get the relation  $C_{l,\pm1}(\boldsymbol{\hat{{y}}})=C_{l,\pm1}(\boldsymbol{\hat{{x}}})e^{\mp\text{i}\pi/2}=\mp\text{i}C_{l,\pm1}(\boldsymbol{\hat{{x}}})$.

	For the discrete rotation angle $\gamma=2\pi/\nu$ ($\nu=3$ for \ch{Fe3GeTe2}), we can consider $\nu$ cases depending on the modulus:
	\begin{equation}
		m\pmod\nu=0,1,...,\nu-1.
	\end{equation} 
	When we perform a rotation of angle $\gamma$ from $x$-axis, the new electric field becomes $\boldsymbol{E}=(\cos{\gamma},\sin{\gamma},0)E$. Because the torkance is invariant under this transformation, we can rewrite the $x$-axis as the new axis and the $\phi$ goes to $\phi-\gamma$. We can apply similar rotation to the $y$-axis, this leads to the following equations
 	\begin{equation}
		\label{eq:rotation_phase}
  		\begin{split}
		&C_{lm}(\boldsymbol{\hat{x}})e^{-\text{i}m\gamma}=C_{lm}(\boldsymbol{\hat{x}})\cos{\gamma}+C_{lm}(\boldsymbol{\hat{y}})\sin{\gamma},\\
            &C_{lm}(\boldsymbol{\hat{y}})e^{-\text{i}m\gamma}=-C_{lm}(\boldsymbol{\hat{x}})\sin{\gamma}+C_{lm}(\boldsymbol{\hat{y}})\cos{\gamma},
	    \end{split}  
        \end{equation}
     where $C_{lm}(\boldsymbol{\hat{x}},(\boldsymbol{\hat{y}}))$ are scalar coefficients that needed to be obtained by fitting the numerical results.      
     The full vector form is $\boldsymbol{C}_{lm}=C_{lm}(\boldsymbol{\hat{x}})\boldsymbol{\hat{x}}+C_{lm}(\boldsymbol{\hat{y}})\boldsymbol{\hat{y}}$, which will contract with the applied Efield vector $\boldsymbol{E}$.  If $m=n\nu$, we see that Eq.~\ref{eq:rotation_phase} cannot be satisfied because the left hand side is always $1$. The reason is that $C_{\nu z}$ symmetry only allows out-of-plane field-induced torque ($\boldsymbol{E}\parallel\boldsymbol{\hat{z}}$) in this case.
     Now let's consider the case $m=n\nu\pm1$, Eq.~\ref{eq:rotation_phase} gives 
     \begin{equation}
		\label{eq:EyandEx}
		C_{l,n\nu\pm1}(\boldsymbol{\hat{y}})=\mp i C_{l,n\nu\pm1}(\boldsymbol{\hat{x}}).
    \end{equation}
    In fact, $m=n\nu\pm1$ are the only two possible cases for $C_{3z}$ rotation symmetry. For $C_{2z}$ symmetry, Eq.~\ref{eq:rotation_phase} is always satisfied for odd $m$. For $C_{4z},C_{6z}$ symmetries, we need to consider more cases, which are summarized in the Table~\ref{table:C4z} and Table~\ref{table:C6z}.
    \begin{table}[htbp]
		\centering
		\begin{tabular}{| c | c| }
			\hline
                $m=4n$&  Not allowed  \\ \hline
			$m=4n\pm 1$ & $\mp i$\\ \hline
			$m=4n+ 2$ & Not allowed \\ \hline
		\end{tabular}
            \caption{The ratio $C_{lm}(\boldsymbol{\hat{y}})/ C_{lm}(\boldsymbol{\hat{x}})$ in $C_{4z}$ systems.}
            \label{table:C4z}
	\end{table}
    \begin{table}[htbp]
		\centering
		\begin{tabular}{| c | c| }
			\hline
                $m=6n$&  Not allowed  \\ \hline
			$m=6n\pm 1$ & $\mp i$\\ \hline
			$m=6n\pm2$ & Not allowed 
                \\ \hline
			$m=6n+3$ & Not allowed
                \\ \hline
		\end{tabular}
            \caption{The ratio $C_{lm}(\boldsymbol{\hat{y}})/ C_{lm}(\boldsymbol{\hat{x}})$ in $C_{6z}$ systems.}
            \label{table:C6z}
	\end{table}     
    
	Now we only need to focus on the applied field in $\boldsymbol{\hat{x}}$ direction to obtain the additional symmetry constraints. Under the mirror reflection respect to $xy$-plane, both the torque $\boldsymbol{T}$ decomposed in $\boldsymbol{\hat{\theta}},\boldsymbol{\hat{\phi}}$ directions and applied field are even. Thus the torkance $\boldsymbol{\tau}$ has to be even under the transformation as well,
	\begin{equation}
		\boldsymbol{\tau}(\theta,\phi)\xrightarrow{\sigma_{xy}}\boldsymbol{\tau}(\theta,\phi+\pi)=e^{im\pi}\boldsymbol{\tau}(\theta,\phi).
	\end{equation}
	This enforces that $m$ must be an even number, i.e., $m=6n\pm2$. The remaining crystal symmetry constraint is due to the $C_{2y}$ rotation symmetry,
	\begin{equation}
		\boldsymbol{\tau}(\theta,\phi)\xrightarrow{C_{2y}}\boldsymbol{\tau}(\pi-\theta,\pi-\phi)=\boldsymbol{\tau}(\pi-\theta,-\phi).
	\end{equation}
	Because $\boldsymbol{T}_{\theta}=\boldsymbol{T}\cdot\boldsymbol{\hat{\theta}},\boldsymbol{T}_{\phi}=\boldsymbol{T}\cdot\boldsymbol{\hat{\phi}}$, and applied field in $\boldsymbol{\hat{x}}$ all flip the sign under $C_{2y}$ rotation, thus the torkance has actually to be even under the rotation.
	To further simplify the constraint, we consider the real and imaginary part of vector spherical harmonics separately by observing the following relation:
	\begin{equation}
		\begin{split}
			&\text{Re}\boldsymbol{Y}^{(\text{D,F})}_{lm}(\pi-\theta,-\phi)=(-1)^{l+m+1} \text{Re}\boldsymbol{Y}^{(\text{D,F})}_{lm}(\theta,\phi), \\
			&\text{Im}\boldsymbol{Y}^{(\text{D,F})}_{lm}(\pi-\theta,-\phi)=(-1)^{l+m} \text{Im}\boldsymbol{Y}^{(\text{D,F})}_{lm}(\theta,\phi).
		\end{split}
	\end{equation}
	Given that $m$ is always even, we are only allowed to have odd/even number $l$ for the real/imaginary part of vector spherical harmonics.
	Last but not the least, we can always decompose the current-induced torque into time-reversal even and odd parts. Under the time-reversal symmetry transformation,
	\begin{equation}
		\boldsymbol{\tau}(\theta,\phi)\xrightarrow{\mathcal{T}}\boldsymbol{\tau}(\pi-\theta,\pi+\phi)=\boldsymbol{\tau}(\pi-\theta,\phi).
	\end{equation}
	The real and imaginary part of vector spherical harmonics both satisfy
	\begin{equation}
		\begin{split}
			&\boldsymbol{Y}^{\rm D}_{lm}(\pi-\theta,\phi)=(-1)^{l+1} \boldsymbol{Y}^{\rm D}_{lm}(\theta,\phi), \\
			&\boldsymbol{Y}^{\rm F}_{lm}(\pi-\theta,\phi)=(-1)^l \boldsymbol{Y}^{\rm F}_{lm}(\theta,\phi). 
		\end{split}
	\end{equation}
	
	The final symmetry-constrained form of time-reversal even torkance under the applied E-field in $\boldsymbol{\hat{x}}$-direction is
	\begin{equation}
		\begin{split}
			\boldsymbol{\tau}^{\text{even}}(\boldsymbol{\hat{x}})=\sum_{lm}&C^{{\rm F}}_{2l,6m\pm2}\text{Im}\boldsymbol{Y}^{\rm F}_{2l,6m\pm2}\\
			+&C^{{\rm D}}_{2l+1,6m\pm2}\text{Re}\boldsymbol{Y}^{\rm D}_{2l+1,6m\pm2}, 
		\end{split} 
		\label{eq:taueven_Ex_full}
	\end{equation}
	and time-reversal odd torkance is
	\begin{equation}
		\begin{split}
			\boldsymbol{\tau}^{\text{odd}}(\boldsymbol{\hat{x}})=\sum_{lm}&C^{{\rm D}}_{2l,6m\pm2}\text{Im}\boldsymbol{Y}^{\rm D}_{2l,6m\pm2}\\
			+&C^{{\rm F}}_{2l+1,6m\pm2}\text{Re}\boldsymbol{Y}^{\rm F}_{2l+1,6m\pm2}.
		\end{split} 
		\label{eq:tauodd_Ex_full}
	\end{equation}
	By utilizing Eq.~\ref{eq:EyandEx}, we can write down the final symmetry-constrained form of torkance under the applied E-field in $\boldsymbol{\hat{y}}$-direction is
	\begin{equation}
		\begin{split}
			\boldsymbol{\tau}^{\text{even}}(\boldsymbol{\hat{y}})=\sum_{lm}&\pm C^{{\rm F}}_{2l,6m\pm2}\text{Re}\boldsymbol{Y}^{\rm F}_{2l,6m\pm2}\\
			&\mp C^{{\rm D}}_{2l+1,6m\pm2}\text{Im}\boldsymbol{Y}^{\rm D}_{2l+1,6m\pm2},  
		\end{split} 
		\label{eq:taueven_Ey_full}
	\end{equation}
	and time-reversal odd torkance is
	\begin{equation}
		\begin{split}
			\boldsymbol{\tau}^{\text{odd}}(\boldsymbol{\hat{y}})=\sum_{lm}&\pm C^{{\rm D}}_{2l,6m\pm2}\text{Re}\boldsymbol{Y}^{\rm D}_{2l,6m\pm2}\\&\mp C^{{\rm F}}_{2l+1,6m\pm2}\text{Im}\boldsymbol{Y}^{\rm F}_{2l+1,6m\pm2}.
		\end{split} 
		\label{eq:tauodd_Ey_full}
	\end{equation}
	Note that scalar coefficients $C_{lm}$ appearing in equations above are the same. We only need to calculate the applied E-field in $\boldsymbol{\hat{x}}$ case and fit the numerical results with the vector spherical harmonics form to obtain the coefficients $C_{lm}$. Note that for this system, changing the direction of the electric field swaps $\text{Re}$ and $\text{Im}$. The differences in these functions correspond to rotations through the azimuth by $\pi/2$.

	\section{Details of the torque calculation.}
	\label{app:first_principles}
	
	The first step is to obtain the tight-binding Hamiltonian in a localized atomic orbital basis using a combination of Quantum Espresso \cite{QE} and Wannier90 \cite{Wannier90}. 
	In the Quantum ESPRESSO implementation, we use the pseudopotentials from PSlibrary \cite{DALCORSO2014337} generated with a fully relativistic calculation using Projector Augmented-Wave method \cite{PAW} and local density approximation exchange correlations \cite{LDA_PZ}. We utilize a $18\times 18 \times 1$ Monkhorst-Pack mesh \cite{MPmesh}, $2~\rm{nm}$ vacuum layer, $2720~{\rm eV}$ cutoff energy, $1.36\times10^{-3}~{\rm eV}$ total energy convergence threshold and obtain the relaxed positions with the forces smaller than $0.02~{\rm eV/nm}$.
	The second step is to use Wannier90 \cite{Wannier90} to obtain the Hamiltonian in an atomic basis.  We project plane-wave solutions onto atomic $s,p,d$ orbitals of Fe atoms, $s,p$ orbitals of Ge and Te atoms without performing maximal localization. We then symmetrize the tight-binding Hamiltonian using TBmodels \cite{TBmodels}. The final symmetrized tight-binding band structures agree very well with these bands from plane-wave methods shown in Fig.~\ref{Fig1:crystal and bands}(c). The band inconsistencies at higher higher above the Fermi level are expected and do not significantly affect our results because states near the Fermi level dominate the torkance calculations through the energy denominator in Eq.~(\ref{eq:eventorkance}).
	
	Equipped with the spin-orbit coupled tight-binding Hamiltonian, We then apply linear response theory to compute the torkance \cite{Freimuth2014,Xue2020SOT,Xue2021SOT_AFM}. We denote the $j^{\rm th}$ component of the torkance in response to an electric field along the $i$-direction with $\tau_{ij}$.  An applied electric field can modify both the electron distribution function and the wavefunction. The linear response from the change of the distribution function is time-reversal odd~\cite{Haney2013} while the linear response from the change of the wavefunction is time-reversal even~\cite{Freimuth2014}. Using the standard Kubo formula~\cite{Freimuth2014,Xue2020SOT}, the even and odd components of the torkance are given by Eq.~(\ref{eq:eventorkance}) and Eq.~(\ref{eq:oddtorkance}) respectively. The even and odd components are also denoted as Fermi sea and Fermi surface terms~\cite{Freimuth2014}. 
	The torque operator is obtained as the change of magnetization with respect to time, 
	\begin{equation}
		\label{eq:torque}
		\boldsymbol{\mathcal{T}}=\frac{d\mathbf{\Delta}}{dt}=\frac{\mathrm{i}}{\hbar}[H,\mathbf{\Delta}]=-\frac{\mathrm{i}}{\hbar}[{\mathbf{\Delta}}\cdot\hat{\mathbf{S}} ,\hat{\mathbf{S}}].
	\end{equation}
	where $\mathbf{S}$ is the spin operator and $\boldsymbol{\Delta}$ is the time-reversal odd spin-dependent exchange-correlation potential.  In order to compute the angular dependence of the torkance, we manually rotate the magnetization direction from the ground state $\theta=0$ to an arbitrary angle $(\theta,\phi)$ by performing a rotation on this time-reversal odd spin-dependent exchange-correlation potential. We use a $80\times160$ mesh of $(\theta,\phi)$ to obtain the angular dependence results shown in Fig.~\ref{Fig:GlobalTorque}.

	We use a very dense ${\bf k}$-mesh of $1200\times1200$ to evaluate the torkance Eqs.~\ref{eq:eventorkance} and \ref{eq:oddtorkance}. Note that we adopt the tight-binding approximation \cite{Souza2019} that Wannier orbitals are perfectly localized on atomic sites and the spin operators $\mathbf{S}$ are described by the Pauli matrices spanned in the Wannier orbital basis in the implementation. 
    We also adopt a constant broadening model to evaluate the longitudinal conductivity~\cite{Freimuth2014},
        \begin{equation}
            \sigma_{xx}=\frac{e^2}{\pi\hbar}\sum_{\boldsymbol{k}n m}\frac{\eta^2{\rm Re}[\bra{\psi_{n\bf{k}}}  \frac{\partial H}{\partial k_x}\ket{\psi_{m\bf{k}}}\bra{\psi_{m\bf{k}}}  \frac{\partial H}{\partial k_x}\ket{\psi_{n\bf{k}}}]}{[(E_m-\mu)^2+\eta^2][(E_n-\mu)^2+\eta^2]}.
            \label{eq:sigma}
        \end{equation}
    Eq.~\ref{eq:sigma} also becomes diverge as a function of $1/\eta$ at the zero broadening limit, similar to Eq.~\ref{eq:oddtorkance}. The sheet resistance then is the reciprocal of longitudinal conductivity per unit cell area.
	
	\bibliography{reference}{}
	\bibliographystyle{apsrev4-2}
\end{document}